\documentstyle[12pt]{article}

\def\matelem#1#2#3{\left<#1\left|#2\vphantom{#1#3}\right|#3\right>}

\begin{document}

\begin{titlepage}
\setcounter{page}{1}

\title{\small\centerline{January 1993 \hfill DOE-ER\,40200-304}
\small\rightline{CPP-50}
\bigskip\bigskip
{\Large\bf Rare $K$-decays as crucial tests\\[2mm] for unified models
          with gauged baryon number}\bigskip}

\author{\normalsize \bf Palash B. Pal\\
\normalsize \em Center for Particle Physics\\
\normalsize \em Physics Department, University of Texas,
              Austin, TX 78712, USA}
\date{}
\maketitle
\vfill
\begin{abstract}
In the grand-unified models based on SU(15)
and SU(16) in which the quarks and leptons are un-unified at the
intermediate stages, we show that
${\rm BR}\; (K_L \to \mu e) \leq 10^{-14}$ and
${\rm BR}\; (K^+ \to \pi^+\mu e) \leq 10^{-14}$ despite the presence
of leptoquark gauge bosons. Thus, the observation of these processes
in the ongoing or upcoming experiments will rule out the models.
\end{abstract}
\vfill
\centerline{PACS numbers: 12.10.Dm, 13.20.Eb, 11.30.Hv}

\thispagestyle{empty}
\end{titlepage}
%
%%%%%%%%%%%%%%%%%%%%%%%%%%%%%%%%%%%%%%%%%%%%%%%%%%%%%%%%%%%%%%%%

Experiments to observe the processes $K_L \to \mu^\pm e^\mp$
and $K^+ \to \pi^+ \mu^\pm e^\mp$ are
underway at Brookhaven \cite{Brex}. When completed, they will probe
a branching ratio as small as $10^{-12}$ for each process.
If the processes are not seen
at that level, it will mean bad news to a lot of theoretical
models beyond the standard model. Here, on the contrary, we point out
a class of grand-unified
models which will definitely be ruled out if the processes
are observed -- not only by the Brookhaven experiment, but with any
branching ratio larger than about $10^{-14}$.

We have in mind the grand unified models based on SU(15)
\cite{Adler89,FrLe90} and SU(16) \cite{PSS75},
which have baryon number as part of its gauge symmetry.\footnote{In
SU(16), lepton number is also part of the gauge symmetry, but we will
not need this for our argument here.}  In the recent literature, there
has been a lot of discussion \cite{Adler89,FrLe90,Pal,BSMS92,DKP}
that such models can have some chains of symmetry breaking where
renormalization group analysis yields very low unification scale, as low
as $10^8$ GeV, without any conflict with the known bounds on proton
lifetime \cite{Pal92,BSMS92,DKP}. It has also the pleasant feature
that the monopole problem vanishes in such models with low unification
scales \cite{Pal91}.

Before proceeding, let us discuss why grand unification models based
on these groups deserve attention. The gauge groups
for these models are not just {\em any} group to play with. SU(15) is
the maximal group for unification for all known fermions in a single
generation, just as SU(16) is if a right handed neutrino is needed
to make the fermion spectrum left-right symmetric \cite{PSS75}.
Just as the minimal grand unification group SU(5) is
interesting for its special status, so are these groups. Secondly,
in all known physics, the fermions transform
as fundamental representations of the non-abelian gauge groups. Quarks
transform as the fundamental representation of the color group SU(3),
left-handed fermions are fundamental representations of the
electroweak SU(2). Thus, it is intriguing to check the idea that all
fermions transform like the fundamental representation of the grand
unified gauge group. Lastly, baryon number and lepton number are known
symmetries of low energy physics. It is interesting to entertain the
possibility that these are gauge symmetries at high energy.

We now begin our argument by briefly describing the SU(15) model.
All known left-chiral fermions of a single
generation transform like the  fundamental representation of
this group:
	\begin{eqnarray}
\Psi_L \equiv \left( u_r u_b u_y \, d_r d_b d_y \;
\hat u _r  \hat u _b \hat u _y \, \hat d _r
\hat d _b \hat d _y \; \nu_e e^- e^+ \right)_L \,.
\label{fermions}
	\end{eqnarray}
The hats denote antiparticles, and $r,b,y$ are three colors. The same
pattern is repeated for other generations.
Mirror fermions are needed to cancel the anomalies. In SU(16), the
only difference is a left-handed antineutrino field which has to be
included in $\Psi_L$ to make it a 16-plet, but this will not affect
our argument.

The interesting chains of symmetry breaking start with the following
pattern:
	\begin{eqnarray}
{\cal G}_0 \stackrel {M_G} {\longrightarrow} {\cal G}_1 \,,
\label{ssb}
	\end{eqnarray}
where ${\cal G}_0$ is the grand unified group, and
	\begin{eqnarray}
{\cal G}_1 &\subseteq& \mbox{SU(12)}_q \times \mbox{SU(3)}_\ell \quad
\mbox{if ${\cal G}_0$ = SU(15)}, \nonumber\\
{\cal G}_1 &\subseteq& \mbox{SU(12)}_q \times \mbox{SU(4)}_\ell \quad
\mbox{if ${\cal G}_0$ = SU(16)} \,.
\label{subgp}
	\end{eqnarray}
Here, the subscript $q$ means that only quarks and antiquarks
transform non-trivially under that part of the gauge group, and the
subscript $\ell$ means the same thing with leptons and antileptons.
This is the crucial element of the models we want to discuss,
which means that although the interactions
of quarks and leptons are unified at the grand-unified level, they are
not so after the first stage of symmetry breaking. However, in the
course of further symmetry breaking, we finally get to a stage where
quarks and leptons transform under the same SU(2)$_L$ and the same
U(1)$_Y$, that is to say that we must get down to the standard model.
But at higher energies, the quarks and leptons may be un-unified, a
possibility which has been discussed by various authors \cite{ununif}
even outside the context of grand-unified models.

	\begin{figure}
\begin{center}
\begin{picture}(90,50)(-10,25)
\thicklines
\put(0,65){\line(1,-1){15}}     \put(7,58){\vector(1,-1){2}}
\put(0,35){\line(1,1){15}}      \put(7,42){\vector(-1,-1){2}}
\put(70,65){\line(-1,-1){15}}   \put(63,58){\vector(1,1){2}}
\put(70,35){\line(-1,1){15}}    \put(63,42){\vector(-1,1){2}}
\multiput(17.5,50)(10,0){4}{\oval(5,5)[t]}
\multiput(22.5,50)(10,0){4}{\oval(5,5)[b]}
\put(5,35){\makebox(0,0){{\large$s_L$}}}
\put(6,65){\makebox(0,0){{\large$\mu^+_L$}}}
\put(65,35){\makebox(0,0){{\large$d_L$}}}
\put(65,65){\makebox(0,0){{\large$e^+_L$}}}
\end{picture}
\end{center}
\caption[]{The process $K_L \to \mu^- e^+$ mediated by gauge bosons.}
\label{f:k2mue}
	\end{figure}
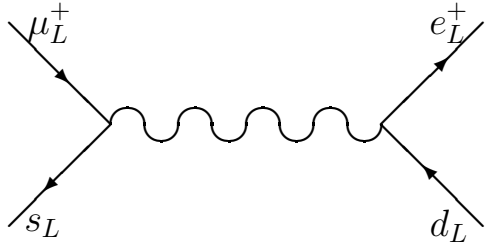
There are gauge bosons in ${\cal G}_0$ which can change any
entry of $\Psi_L$ into any other. Thus,
there will be tree-level diagrams, mediated by gauge boson exchange,
which contributes to the process $K_L \to \mu^-e^+$. One example is shown
in Fig. \ref{f:k2mue}, which is the process $\mu^+_Ld_L \to e^+_Ls_L$
at the fundamental level. Since the gauge bosons responsible for these
decays are leptoquark gauge bosons
outside the SU(12)$_q$ subgroup, their masses are of order
$M_G$, the unification scale. Thus, from Fig. \ref{f:k2mue}, one
obtains a transition operator as follows:
	\begin{eqnarray}
{\cal O}_1 &\simeq &
{g^2 \over M_G^2} \; [ \overline s_L \gamma^\lambda {\mu^+_L} ]
\, [\overline {e^+_L} \gamma_\lambda d_L] \nonumber\\
&=& -\, {g^2 \over M_G^2} \; [ \overline s_L \gamma^\lambda d_L ]
\, [\overline {\mu_R} \gamma_\lambda e_R] \,,
\label{O1}
	\end{eqnarray}
where the last form is obtained by applying Fierz transformation
and then using the identity $\overline {e^+_L} \gamma_\lambda {\mu^+_L}
= -\overline {\mu_R} \gamma_\lambda e_R$. Similarly,
there is another diagram which, at the fundamental level, induces the
process $e^-_L\hat s_L \to \mu^-_L\hat d_L$. This gives an operator
	\begin{eqnarray}
{\cal O}_2 &\simeq &
-\, {g^2 \over M_G^2} \; [ \overline s_R \gamma^\lambda d_R]
\, [\overline {\mu_L} \gamma_\lambda e_L] \,,
\label{O2}
	\end{eqnarray}
Adding the two and using definition of the kaon decay constant $f_K$,
we obtain a transition operator for $K^0 \to \mu^-e^+$. Similarly, one
can determine the operator for $\overline{K^0}\to \mu^-e^+$. Taking
the superposition, we finally obtain the matrix element for the
transition $K_L\to \mu^- e^+$:
	\begin{eqnarray}
{\cal A} (K_L \to \mu e) &\simeq & {ig^2 f_K m_\mu \over
\surd 2M_G^2} \, [\overline \mu \gamma_5 e] \,.
\label{A1}
	\end{eqnarray}
This gives a decay rate
	\begin{eqnarray}
\Gamma (K_L \to \mu e)_G &\simeq & {m_K \over 16\pi} {g^4 \over M_G^4} \,
f_K^2 m_\mu^2 \left( 1- {m_\mu^2 \over m_K^2} \right)^2 \,,
\label{Gam1}
	\end{eqnarray}
where the subscript $G$ reminds us that it is the contribution
from gauge boson exchange. The only unknown parameter in this
expression is $M_G$. In general, if $K_L \to \mu e$ is mediated by
leptoquark gauge bosons, the mass of these gauge bosons are not
constrained by phenomenological arguments, so that the branching ratio
can be large. However, we now show that in the class of models
we are considering, $M_G$ has a lower bound, which implies
an upper bound on the decay rate given above.

	\begin{figure}
\begin{center}
\begin{picture}(90,50)(-10,25)
\thicklines
\put(0,65){\line(1,-1){15}}     \put(7,58){\vector(1,-1){2}}
\put(0,35){\line(1,1){15}}      \put(7,42){\vector(-1,-1){2}}
\put(70,65){\line(-1,-1){15}}   \put(63,58){\vector(1,1){2}}
\put(70,35){\line(-1,1){15}}    \put(63,42){\vector(-1,1){2}}
\multiput(17.5,50)(10,0){4}{\oval(5,5)[t]}
\multiput(22.5,50)(10,0){4}{\oval(5,5)[b]}
\put(5,35){\makebox(0,0){{\large$s$}}}
\put(6,65){\makebox(0,0){{\large$\widehat s$}}}
\put(65,35){\makebox(0,0){{\large$d$}}}
\put(65,65){\makebox(0,0){{\large$\widehat d$}}}
\end{picture}
\end{center}
\caption[]{Contribution to $K^0$-$\overline{K^0}$ from gauge boson
exchange at the tree level.}
\label{f:klks}
	\end{figure}
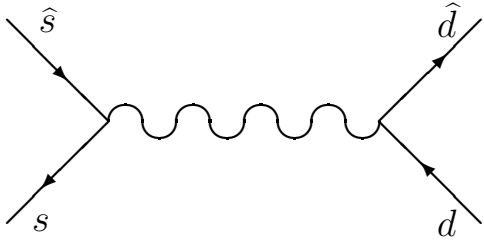
To see this, we consider the diagram of Fig. \ref{f:klks}, which gives
a tree-level contribution to the $K^0$-$\overline{K^0}$ amplitude
\cite{Pal}.
Note that the gauge bosons present in this diagram are those contained
in the SU(12)$_q$ subgroup shown in Eq (\ref{subgp}), and therefore
their masses, $M_B$, must be less than or equal to $M_G$.
Then, Fig. \ref{f:klks} gives
	\begin{eqnarray}
{\cal A} (K^0 \mbox{-} \overline{K^0}) &\simeq &{g^2 \over M_B^2} \,
[\overline d_L^b  \gamma_\lambda \hat d_L^a]
[\overline{\hat{s}}_L^a  \gamma^\lambda s_L^b] \nonumber\\
&=&  - \, {g^2 \over M_B^2} \, [\overline d_L^b  \gamma^\lambda s_L^b]
[\overline d_R^a \gamma_\lambda s_R^a] \,,
\label{Ak0-k0^}
	\end{eqnarray}
where $a$ and $b$ are summed color indices. We now define the matrix element
	\begin{eqnarray}
\matelem {K^0} {[\overline{d}_L^b  \gamma^\lambda s_L^b]
[\overline{d}_R^a \gamma_\lambda s_R^a]}
{\overline{K^0}} \equiv {\cal B} f_K^2 m_K \,,
\label{defb}
	\end{eqnarray}
where $\cal B$ is a parameter estimated later.
The contribution of the operator in Eq (\ref{Ak0-k0^}) to the
$K_L$-$K_S$ mass difference is given by
	\begin{eqnarray}
\left. \vphantom{{g^2 \over M_B^2}}
\Delta m_K \right|_{\rm Fig.\ref{f:klks}}
= {g^2 \over M_B^2} \, {\cal B} f_K^2 m_K \,.
\label{DmK}
	\end{eqnarray}
Since the standard model contribution gives roughly the right
magnitude and right sign for the experimentally measured value
of $\Delta m_K = 3.5  \times 10^{-15}$ GeV, we can say that the
magnitude of the contribution from  Eq (\ref{DmK}) is less than the
experimental value. Putting the vacuum insertion estimate \cite{BBS}
${\cal B}=2.6$, we thus predict
	\begin{eqnarray}
M_B/g \geq 2.3 \times 10^6\, {\rm GeV}. \,,
\label{MBbd}
	\end{eqnarray}
using $f_K = 114$ MeV and $m_K= 494$ MeV.

This puts a lower bound on the $K_L \to \mu e$ decay rate since, as
mentioned earlier, $M_G \geq M_B$. Thus, we can use Eq (\ref{Gam1}) to
write
	\begin{eqnarray}
\Gamma (K_L \to \mu e)_G &\leq & {m_K \over 16\pi} {g^4 \over M_B^4} \,
f_K^2 m_\mu^2 \left( 1- {m_\mu^2 \over m_K^2} \right)^2 \,.
\label{Gam1bd}
	\end{eqnarray}
Using Eq (\ref{MBbd}) and $\Gamma(K_L \to \mbox{all}) =
1.3\times 10^{-17}$ GeV, we obtain
	\begin{eqnarray}
\mbox{BR} \, (K_L \to \mu e)_G &\leq & 4\times 10^{-15} \,.
\label{finalbd}
	\end{eqnarray}
It is of course
true that because of renormalization effects, the gauge coupling
constants appearing in Eqs (\ref{Gam1}) and (\ref{DmK}) need not be
the same, but the correction cannot be large enough to bring the
branching ratio above $10^{-12}$, which is the limit sought for in the
experiment. Notice that the result only depends on the fact that
the unified group breaks in a single step to ${\cal G}_1$ given by
Eq (\ref{subgp}), and is independent of any subsequent symmetry breaking.

It is possible to derive similar conclusions regarding the decay of
charged kaons which violate muon number. Consider the process
$K^+ \to \pi^+\mu^-e^+$, for example. At the quark level, the
operators responsible for this process are those given in Eqs (\ref{O1})
and (\ref{O2}). Taking the hadronic matrix elements and adding up the
two contributions, we obtain
	\begin{eqnarray}
{\cal A} (K^+ \to \pi^+\mu^-e^+) = {g^2 \over M_G^2} [
(k+p)_\lambda f_+ + (k-p)_\lambda f_- ] \;
[\overline {\mu} \gamma^\lambda e ] \,,
	\end{eqnarray}
where $k$ and $p$ are the kaon and pion momenta, and
the form factors $f_\pm$ are functions of $(k-p)^2$,
known from $K_{e3}$ and $K_{\mu 3}$
decays: $f_+(q^2) = 1+ 0.032q^2/m_\pi^2$ and $f_-=-0.322$.
Neglecting the masses of the decay particles, this gives a
decay rate
	\begin{eqnarray}
\Gamma (K^+ \to \pi^+\mu e) _G = {m_K^5 \over 768\pi^3} \, {g^4
\over M_G^4} \left< f_+^2 \right> \,,
\label{K+rate}
	\end{eqnarray}
where $\left< f_+^2 \right>$ is an energy-averaged value of $f_+^2$.
Replacing this average by the maximum possible value of $f_+^2$, which
is 1.44, and using $M_G\geq M_B$, we can use Eq (\ref{MBbd}) to find
an upper limit on the rate in Eq (\ref{K+rate}). Using
$\Gamma(K^+ \to \mbox{all}) = 5.3 \times 10^{-17}$ GeV, we obtain
	\begin{eqnarray}
\mbox {BR} (K^+ \to \pi^+\mu^-e^+) _G \leq 8\times 10^{-16} \,,
\label{finalbd2}
	\end{eqnarray}
which is once again much smaller than the value sought for in the
experiment. Thus, observation of either $K_L \to \mu e$ or
$K^+ \to \pi^+\mu^-e^+$ will be a death-knell to these wide class of
models.

One may wonder whether the powerful results given above
are marred by Higgs boson
exchange. We show that they are not. For this, consider Higgs
bosons replacing gauge bosons in Fig. \ref{f:k2mue} (with the associated
changes in chirality on fermion lines). A straightforward comparison
of this diagram with the gauge boson mediated diagram of Fig.
\ref{f:k2mue} shows that
	\begin{eqnarray}
{\Gamma (K_L \to \mu e)_H \over \Gamma (K_L \to \mu e)_G}
\simeq {h_1^2 h_2^2 \over g^4} \, {M_H^4 \over M_G^4}
\simeq 10^{-12} {M_H^4 \over M_G^4} \,,
	\end{eqnarray}
where $h_1$ and $h_2$ are Yukawa couplings pertaining to the first and
the second generations, whose values should be around $10^{-4}$ and
$10^{-3}$ respectively if we ignore fermion mixing. Using Eq
(\ref{finalbd}), we thus see that the Higgs mediated contribution can
give a branching ratio larger than $10^{-14}$ only if $M_H^2/M_G^2 < 6
\times 10^{-7}$. However, these are colored Higgs bosons we are
talking about, whose masses should be of order the unification mass.
Thus, we conclude that the contribution from these colored Higgs
bosons are negligible. The same result applies for charged kaon decays
into $\pi\mu e$.

Next, we consider the possibility of exchange of flavor-changing
neutral Higgs (FCNH) bosons, which may have couplings like $\bar dsH$ and
$\bar \mu eH$, and therefore can mediate flavor changing processes.
However, the important point to realize is that there is no such Higgs
boson in the model. The reason is that the only Higgs bosons which
couple to the fermions are the ones which transform like a rank-2
tensor representation of the gauge group. Thus, let us consider the
case when the model contains a Higgs boson multiplet $S^{\{ij\}}$,
where the curly bracket denotes symmetric indices. In this case, the
masses of the up-type quarks are derived only from the vacuum
expectation value (VEV) of the components $S^{\{u\hat u\}}$, which
means the color singlet combination of the components whose one index
has the same quantum number of $u$ and the other of $\hat u$.
Similarly, the down-type quarks obtain masses only from the VEV of
$S^{\{d\hat d\}}$, and the charged lepton from $S^{\{e^-e^+\}}$. In
the minimal model, there is one VEV of each kind,
and so the model is free from FCNH interactions\footnote{One can
always contemplate models with many copies of the same Higgs boson
representations, each with its own VEV, in which case FCNH will exist.
What we emphasize here is that this need not happen in these models,
as opposed to, say SO(10) models where FCNH must exist.}
\cite{GlWe77}. Thus, the bounds given in Eqs
(\ref{finalbd}) and (\ref{finalbd2}) are not disturbed.

In summary, we have shown that the branching ratio for $K_L \to
\mu^-e^+$ and $K^+ \to \pi^+\mu^-e^+$
cannot be larger than $10^{-14}$ in a large class of models. This
result is obtained despite the presence of leptoquark gauge bosons in
the model. The reason is that, there are diquark gauge bosons which
are lighter than the leptoquarks, and their mass is constrained from
the $K_L$-$K_S$ mass difference.
The same comments applies for the charge-conjugate
modes $K_L \to \mu^+e^-$ and $K^+ \to \pi^+\mu^+e^-$.
Observation of these modes in
the ongoing or upcoming experiments should then rule out these models.

The author is indebted to K. Lang for discussions about the undergoing
Brookhaven experiment, and to B. Lynn for encouragement for emphasizing
the constraints on $K_L\to \mu e$ in SU(15) models.


\newpage
				\begin{thebibliography}{[WW]}

\bibitem{Brex} Brookhaven experiments E791 and E865.

\bibitem{Adler89} S.~L. Adler: Phys. Lett. {\bf B225}, 143 (1989).

\bibitem{FrLe90} P.~H. Frampton and B-H. Lee: Phys. Rev. Lett.
{\bf 64}, 619 (1990).

\bibitem {PSS75} J. C. Pati, A. Salam and J. Strathdee: Nuovo Cimento
{\bf 26A}, 77 (1975); Nucl. Phys. {\bf B185}, 445 (1981).

\bibitem{Pal} P.~B. Pal: Phys. Rev. {\bf D43}, 236 (1991).

\bibitem{BSMS92} B. Brahmachari, U. Sarkar, R.~B. Mann and T.~G.
Steele: Phys. Rev. {\bf D45}, 2467 (1992).

\bibitem{DKP} N.~G. Deshpande, E. Keith and P.~B. Pal: University of
Oregon preprint OITS-493 (August 1992), to appear in Phys. Rev. D.

\bibitem{Pal92} P.~B. Pal: Phys. Rev. {\bf D45}, 2566 (1992).

\bibitem{Pal91} P.~B. Pal: Phys. Rev. {\bf D44}, R1366 (1991).

\bibitem{ununif} S. Rajpoot: Mod. Phys. Lett. {\bf 1}, 645 (1986);
H. Georgi, E. Jenkins and E.~H. Simmons:
Phys. Rev. Lett. {\bf 62}, 2789 (1989); {\bf 63}, 1540 (E)
(1989); D. Choudhury: Mod. Phys. Lett. {\bf A6}, 1185 (1991).

\bibitem{BBS} G. Beall, M. Bander and A. Soni: Phys. Rev. Lett. {\bf
48}, 848 (1984).

\bibitem{GlWe77} S. L. Glashow and S. Weinberg: Phys. Rev. D {\bf 15},
1958 (1977).

				\end{thebibliography}
\end{document}